\documentclass[12pt]{article}
\usepackage{geometry}	
\usepackage{graphicx} 
\usepackage{epstopdf}
\usepackage{url}	
\usepackage{setspace}\spacing{1.213}	
\usepackage[bottom]{footmisc}			
\usepackage{rotating,float}		
\usepackage[justification=centering,font=small]{caption} 
\usepackage{verbatim}  
\usepackage{adjustbox} 
\usepackage{relsize}   
\usepackage{amsmath,amssymb}\allowdisplaybreaks[1] 

\usepackage{appendix} 
\usepackage{placeins}
\usepackage[small,compact]{titlesec}  
\usepackage[bookmarks=true]{hyperref} 
\usepackage{bookmark}
\usepackage{pdflscape}
\usepackage{url}
\usepackage{ragged2e}

\usepackage{xcolor}	
\hypersetup{
	colorlinks,
	linkcolor={red!50!black},
	citecolor={blue!50!black},
	urlcolor={blue!80!black}
}

\usepackage[authoryear,round]{natbib}

\usepackage{amsthm}
\theoremstyle{plain}

\theoremstyle{definition}

\newtheoremstyle{indenteddefinition}
	{}
	{}
	{\hangindent=2em}
	{}
	{\bfseries}
	{.}
	{.5em}
	{}
\theoremstyle{indenteddefinition}

\newcounter{assumptiongroup}\stepcounter{assumptiongroup}

\title{A Model of a Randomized Experiment with an Application to the PROWESS Clinical Trial\thanks{This manuscript was first circulated as NBER Working Paper 25670 in March of 2019. I thank Neil Christy, Tory Do, Bailey Flanigan, Pauline Mourot, Srajal Nayak, and Matthew Tauzer for excellent research assistance.  I also thank Charles Antonelli,  Bennett Fauber, and Advanced Research Computing at the University of Michigan, as well as Andrew Sherman and the Yale University Faculty of Arts and Sciences High Performance Computing Center. David Wilson provided helpful comments.}}
\author{Amanda E. Kowalski}
\date{March 13, 2019}

\begin{document}
	\maketitle
	
\begin{center}
This paper has been combined with ``Counting Defiers'' (\url{https://arxiv.org/abs/1908.05811}) and superseded by ``Counting Defiers: Examples from Health Care'' (\url{https://arxiv.org/abs/1912.06739}) as of

July 17, 2020
\end{center}
\bigskip	
	
	\begin{abstract}
	I develop a model of a randomized experiment with a binary intervention and a binary outcome. Potential outcomes in the intervention and control groups give rise to four types of participants.  Fixing ideas such that the outcome is mortality, some participants would live regardless, others would be saved, others would be killed, and others would die regardless. These potential outcome types are not observable. However, I use the model to develop estimators of the number of participants of each type.  The model relies on the randomization within the experiment and on deductive reasoning. I apply the model to an important clinical trial, the PROWESS trial, and I perform a Monte Carlo simulation calibrated to estimates from the trial.  The reduced form from the trial shows a reduction in mortality, which provided a rationale for FDA approval. However, I find that the intervention killed two participants for every three it saved.	
	
	\end{abstract}

	\pagebreak

\section{Introduction}

Consider a randomized experiment with a binary intervention and a binary outcome.  To fix ideas, suppose that the outcome is mortality.  At the end of the experiment, we observe whether each participant is alive or dead.  We also observe whether each participant was randomized into the intervention group or the control group.  Using these data, we can calculate the reduced form, which is equal to the fraction of participants observed dead in the intervention group minus the fraction of participants observed dead in the control group.  If the reduced form is negative, then we conclude that, on average, the intervention reduced mortality.  If the intervention was access to a medical treatment, then the reduced form provides a rationale to expand access to the treatment.  

However, there is a possibility that even though the intervention reduces mortality on average, some individuals would be killed by it.  To be precise, within the experiment, the participants who would be killed are those that would die if they were assigned to the intervention group but would live if they were assigned to the control group.  If we could identify such individuals, then their prevalence might provide a rationale against expanding access to the treatment.  

However, individuals who would be killed are difficult to identify on the individual level because it is not possible to observe what mortality would have been in the control group for a participant assigned to the intervention group and vice versa.  This problem is well-known, and it has been formalized with the concept of ``potential outcomes"  \citep{rubin1974,rubin1977,holland1986}. In the context of the experiment, each participant has a potential outcome in the control group and a potential outcome in the intervention group, but only one potential outcome is observed for each participant.  On the basis of all combinations of potential outcomes, there are four types of participants in the experiment.  In addition to those who would be killed, there are those who would be saved, those who would live regardless, and those who would die regardless.    

The reduced form gives the difference between the fraction of participants who would be killed and the fraction of participants who would be saved.  To see this, recall that the reduced form is equal to the fraction of participants observed dead in the intervention group minus the fraction of participants observed dead in the control group.  In terms of potential outcome types, the fraction of participants observed dead in the intervention group is equal to the fraction who would be killed plus the fraction who would die regardless.  (By definition, those who would be saved and those who would live regardless would not contribute to the fraction observed dead in the intervention group.)  By similar reasoning, the fraction of participants observed dead in the control group is equal to the fraction who would be saved plus the fraction who would die regardless.  Because the randomization assures that the expected fraction who would die regardless is the same in the intervention and control groups, the reduced form is equal to the difference between the fraction of participants who would be killed and the fraction of participants who would be saved. 

Beyond the \emph{difference} between the fraction of participants who would be killed and the fraction of participants who would be saved, it could also be useful to know the ratio of the two fractions and the absolute numbers of participants saved and killed.  For example, in moral philosophy, the ``trolley problem'' and the related ``transplant problem," involve whether it is ethical to kill one person to save five others, either by diverting a trolley or by reallocating organs  \citep{foot1967,thomson1985}. Randomized experiments no doubt involve ethical questions that are difficult to answer, and it is perhaps even more difficult to answer those questions without the numbers of participants who would be saved and killed.    
   
I develop estimators of the numbers of participants of each of the four potential outcome types: those who would be saved, those who would be killed, those who would die regardless, and those who would live regardless.  I begin by developing a model of a randomized experiment.  The model relies on the randomization itself and on deductive reasoning.  Using the model, I derive an expression for the probability of the data that I observe as a function of the number of participants of each potential outcome type and the intended fraction of participants in the intervention group.  I then propose two estimators of the number of participants of each potential outcome type.  The first is a maximum likelihood estimator.  It maximizes the likelihood of the number of participants of each potential outcome type given the data that I observe and the intended fraction of participants in the intervention group.  The maximum likelihood estimator is not yet computationally tractable via a variety of approaches, so I develop a second estimator.  The second estimator is a least squares estimator that minimizes a weighted function of the randomization error within and across potential outcome types, and it is computationally tractable.

I apply the least squares estimator to data from an important clinical trial, the Recombinant Human Activated Protein C Worldwide Evaluation of Severe Sepsis (PROWESS) clinical trial \citep{bernard2001}.  This trial randomized access to an intravenously-administered biologic drug referred to as recombinant human activated protein C, drotrecogin alfa activated, or Xigris (manufactured by Eli Lilly).  Trial participants included 1690 patients with severe sepsis, a life-threatening condition.  The reduced form for 28-day mortality showed that the intervention reduced mortality by 6 percentage points, which was sizable relative to the 31\% 28-day mortality rate in the  control group.  On the basis of this result, the trial suspended enrollment, and the FDA expedited approval of the drug for patients with severe sepsis in 2001.  

However, the FDA approval was controversial in part because an alternative reduced form within the trial showed that the intervention increased serious bleeding by 1.5 percentage points \citep{siegel2002,warren2002}.  The increase in serious bleeding provides a potential mechanism through which the intervention could kill participants.  The article that presented the trial results claimed that ``1 additional life would be saved for every 16 patients" randomized into the intervention group \citep{bernard2001}. Presumably, the origin of this claim is that 1/16 is approximately 6 percentage points, the size of the reduced form.  However, given the relationship between the reduced form and the number of patients saved and killed, it would be more accurate to say that 1 additional life would be saved \emph{on net} for every 16 patients randomized into the intervention group.  The reduced form indicates that the intervention saved 6 percentage points more participants than it killed, but the number of participants killed, if any, is not observable, and neither is the ratio of participants killed to saved.  The model paves the way for me to estimate those statistics.

To apply the model, I need very limited aggregate data, which I take directly from the article that presented the trial results \citep{bernard2001}.  Specifically, the data include the number of participants who died in the intervention group, the number of participants who lived in the intervention group, the number of participants who died in the control group, and the number of participants who lived in the control group.  I also take as given the intended fraction of participants in the intervention group, which was reported to be one half.

Within the PROWESS trial, I find that 57\% of participants would live regardless, 18\% would be saved, 12\% would be killed, and 13\% would die regardless.  Putting these numbers together, the intervention would have no effect on 70\% of participants, but among the remaining participants, it would kill two participants for every three it saved (12\%/18\%=2/3).  I find that the total number of participants who would be killed by the intervention is 205. However, about half of those participants were were randomized into the control group, so they were not killed.  The intermediate output of my estimator shows that 103 participants were killed within the trial.   These results raise ethical questions about whether to expand access to the drug.  It is notable that based on subsequent evidence, the manufacturer, Eli Lilly, voluntarily withdrew the drug from the market worldwide in 2011 \citep{fda2011}.

To assess the performance of my estimator, I perform a Monte Carlo exercise calibrated to the PROWESS trial.  Because the experiment and the Monte Carlo exercise both involve randomization, the Monte Carlo exercise provides an intuitive illustration of the model. I begin by generating experiments of the same size as the PROWESS trial.  In each experiment, I calibrate the true number of participants of each potential outcome type to my estimates from the trial, and I generate the data.  I then apply my estimator to recover new estimates in each experiment.  Across simulations, I find that the average bias in the estimated number of participants of each potential outcome type is fairly stable across potential outcome types at approximately 42 participants, which represents 2.5\% of the full sample size.  The RMSE in the estimate of each potential outcome type is approximately 7.3\% of the full sample size.  

In the next section, I introduce the model, which I use to derive the probability of the observed data.  I present the maximum likelihood estimator and the least squares estimator in Section~\ref{sec:estimation}.  I present the empirical application to the PROWESS trial and the related Monte Carlo simulation in Sections~ \ref{sec:empirical} and \ref{sec:montecarlo}.  I conclude in Section~\ref{sec:conclusion}.

\section{Model} \label{sec:model}

Consider a participant with potential outcome type $i$. For example, the participant could be of the type that would live regardless of the intervention.  He does not know his potential outcome type.  He does not even know the distribution of potential outcome types, so he is willing to be randomized into the intervention group or the control group.  

At the start of the experiment, the experimenter has already determined the intended fraction of participants in the intervention group $p$.  If  $p=1/2$, then the experimenter flips a fair coin to determine whether the participant will be randomly assigned to the intervention or control group. If $p \neq 1/2$, then the experimenter uses a different randomization method.  Either way, the indicator for whether the participant is assigned to the intervention group is a random variable that is distributed according to a Bernoulli distribution with probability of success $p$.  

Suppose that there are other participants of the same potential outcome type as the participant in question.  Denote the total number of participants of potential outcome type $i$ as $t(i)$.  Randomization occurs independently for each participant with the same probability of assignment to the intervention group $p$.  Therefore, the total number of participants of potential outcome type $i$ who are randomly assigned to the intervention group is equal to the sum of independent and identically distributed Bernoulli random variables.  The sum of independent and identically distributed Bernoulli random variables has a binomial distribution with parameters that represent the number of trials and the probability of success in each trial.  Therefore, the total number of participants of potential outcome type $i$ who are randomly assigned to the intervention group is distributed according to a binomial distribution with parameters $t(i)$ and $p$.

\begin{figure}[!hbt]
	\caption{Matrix that Relates Potential Outcome Types and Observed Outcome Groups}
	\centering
	\includegraphics[width=1\linewidth]{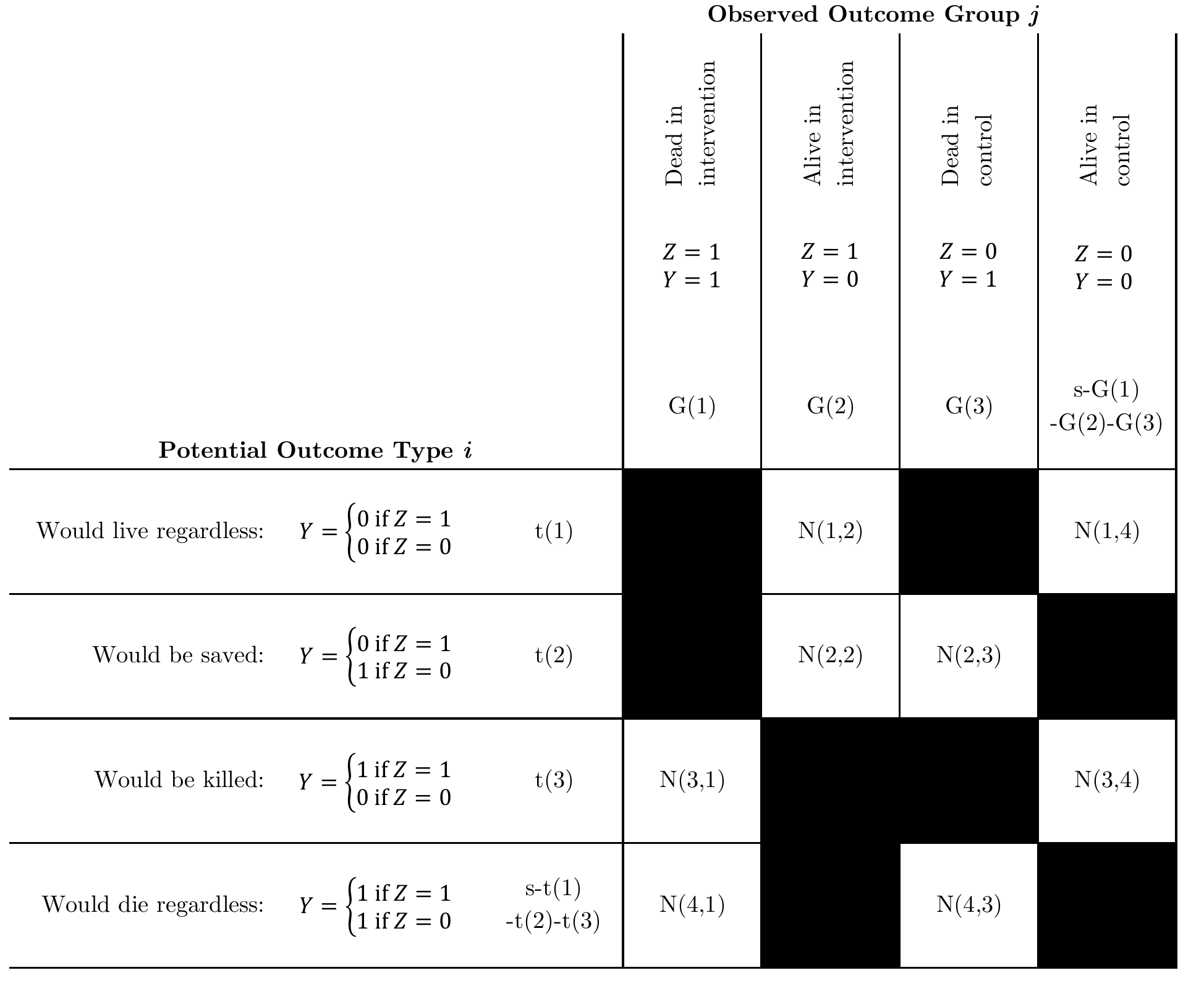}
	\label{fig:expMatrix} \\
	\vspace{-40pt}
	\begin{minipage}{1\linewidth}
	\scriptsize
	\emph{Note}. $Y$ represents mortality, and $Z$ represents assignment to the intervention group.  $N(i,j)$, the number of participants of potential outcome type $i$ in observed outcome group $j$, must be equal to zero in all shaded cells.  The intended fraction of participants in the intervention group is $p$.
	\end{minipage}
\end{figure}

There are four possible potential outcome types.  The matrix in Figure~\ref{fig:expMatrix} includes a separate row for each potential outcome type $i$.  The first row includes the $t(1)$ participants who would live regardless of the intervention ($Y=0$ if $Z=1$ and $Y=0$ if $Z=0$); the second row includes the $t(2)$ participants who would be saved by the intervention ($Y=0$ if $Z=1$ and $Y=1$ if $Z=0$); the third row includes the $t(3)$ participants who would be killed by the intervention ($Y=1$ if $Z=1$ and $Y=0$ if $Z=0$); and the fourth row includes the $t(4)$ participants who would die regardless of the intervention ($Y=1$ if $Z=1$ and $Y=1$ if $Z=0$).  

Importantly, there is effectively a separate randomized experiment within each potential outcome outcome type $i$.  Therefore, the total number of participants of any potential outcome type who are randomly assigned to the intervention group is independent of the total number of participants of any other potential outcome type who are randomly assigned to the intervention group.   This independence proves useful in expressing the distribution of the data as a function of independent binomial random variables. 

The data consist of the total number of participants observed in each of four possible observed outcome groups at the end of the experiment, after the intervention group has received access to the treatment.  The matrix in Figure~\ref{fig:expMatrix} includes a separate column for each observed outcome group $j$.  The first column includes the $G(1)$ participants observed dead in the intervention group ($Z=1$ and $Y=1$); the second column includes the $G(2)$ participants observed alive in the intervention group ($Z=1$ and $Y=0$); the third column includes the $G(3)$ participants observed dead in the control group ($Z=0$ and $Y=1$); and the fourth column includes the $G(4)$ participants observed alive in the control group ($Z=0$ and $Y=0$). Note that $G(j)$ is a random variable, which is why I denote it with a capital letter. The data consist of the vector of $G(1)$, $G(2)$, $G(3)$, and $G(4)$, which I represent with $\mathbf{G}$, in bold to indicate that it is a vector. In what follows, I denote a realization of $G(i)$ with $g(i)$ and a realization of $\mathbf{G}$ with $\mathbf{g}$.  I also denote the vector of $t(1)$, $t(2)$, $t(3)$, and $t(4)$ with $\mathbf{t}$.  

The purpose of my model is to yield an expression for $P(\mathbf{G}=\mathbf{g} \mid \mathbf{t},p)$, the probability of the data vector $\mathbf{G}$ in terms of the vector of potential outcome types $\mathbf{t}$ and the intended fraction of participants in the intervention group $p$.  Toward that end, the matrix in Figure~\ref{fig:expMatrix} relates the four potential outcome types to the four observed outcome groups.  Each cell in the matrix includes the $N(i,j)$ participants with potential outcome type $i$ in observed outcome group $j$. In what follows, I denote the matrix of all $N(i,j)$ with $\mathbf{N}$, in bold to indicate that it is a matrix. The matrix $\mathbf{N}$ has 16 total cells.

However, 8 of the cells cannot have any participants in them.  Consider the upper left cell, which includes participants of the potential outcome type $i=1$ and observed outcome group $j=1$.  Participants of potential outcome type $i=1$ would live regardless of assignment to the intervention group, and participants of observed outcome group $j=1$ are observed dead in the intervention group.  By definition, participants of the type who would live regardless of the assignment to the intervention group will not be observed dead in the intervention group, so it must be the case that $N(1,1)=0$. Similarly, $N(1,3)=0$ participants of the type $i=1$ who would live regardless of assignment to the intervention group will not be observed dead in the control group $j=3$.  The logic for other cells proceeds similarly. I shade all 8 cells that cannot have any participants in them. 

By deductive reasoning  based on the shaded cells, I can express the distribution of the data as follows: 
\begin{align}
P(\mathbf{G}=\mathbf{g} \mid \mathbf{t},p) &= P\bigg(G(1)=g(1), G(2)=g(2),G(3)=g(3),G(4)=g(4) \mid \mathbf{t},p\bigg) \label{eq:start} \\
&= P\bigg(N(3,1)+N(4,1)=g(1), \nonumber \\ 
&\qquad\quad N(1,2)+N(2,2)=g(2), \nonumber \\
&\qquad\quad N(2,3)+N(4,3)=g(3),  \nonumber \\
&\qquad\quad N(1,4)+N(3,4)=g(4) \mid \mathbf{t},p\bigg) \label{eq:eight} \\
&= P\bigg(N(3,1)+N(4,1)=g(1), \nonumber \\
&\qquad\quad N(1,2)+N(2,2)=g(2), \nonumber \\
&\qquad\quad N(2,2)+N(4,1)=t(2)+t(4)-g(3), \nonumber \\
&\qquad\quad N(1,2)+N(3,1)=t(1)+t(3)-g(4) \mid \mathbf{t},p\bigg). \label{eq:four}
\end{align}
I transition from (\ref{eq:start}) to (\ref{eq:eight}) by expressing each $G(j)$ as the sum of the nonzero participant counts in column $j$.  The resulting expression is in terms of the participant counts in all 8 cells that are not shaded in Figure~\ref{fig:expMatrix}.  I simplify the expression further by recognizing that within any potential outcome type $i$, the number of participants in the control group is equal to $t(i)$ minus the number of participants in the intervention group.  Therefore, I can express (\ref{eq:four}) in terms of only four random variables: $N(1,2)$, $N(2,2)$, $N(3,1)$, and $N(4,1)$.  These random variables represent the realized number of participants assigned to the intervention group in each of the four potential outcome types.  

From above, we know that the number of participants assigned to the intervention group in each potential outcome type is a binomial random variable with parameters $t(i)$ and $p$.  Furthermore, we know that the number of participants assigned to the intervention group is independent for each type.  Therefore, (\ref{eq:four}) is a convolution of four independent probabilities with known distributions.  I aim to express (\ref{eq:four}) as the product of four independent probabilities with known distributions. 
  
Using the approach to deconvolution for discrete random variables, I express (\ref{eq:four}) as the sum of the probability of each possible realization $\ell$ of $N(1,2)$.  I then rearrange terms to obtain an expression in terms of the joint probability of $N(1,2)$, $N(2,2)$, $N(3,1)$, and $N(4,1)$:

\begin{align*}
P(\mathbf{G}=\mathbf{g} \mid \mathbf{t},p) &= \sum_{\ell=0}^{t(1)} P\bigg(N(1,2)=\ell, \\
&\qquad\qquad\;\ N(3,1) + N(4,1)=g(1), \\
&\qquad\qquad\;\ \ell+N(2,2)=g(2), \\
&\qquad\qquad\;\ N(2,2)+N(4,1)=t(2)+t(4)-g(3), \\
&\qquad\qquad\;\ \ell+N(3,4)=t(1)+t(3)-g(4) \mid \mathbf{t},p\bigg) \\
&= \sum_{\ell=0}^{t(1)} P\bigg(N(1,2)=\ell, \\
&\qquad\qquad\;\ N(2,2)=g(2)-\ell, \\
&\qquad\qquad\;\ N(3,1)=t(1)+t(3)-g(4)-\ell, \\
&\qquad\qquad\;\ N(4,1)=g(1)+g(4)+\ell-t(1)-t(3), \\
&\qquad\qquad\;\ \sum_{j=1}^4 g(j)=\sum_{i=1}^4 t(i) \mid \mathbf{t},p\bigg).
\end{align*}
Since $N(1,2)$, $N(2,2)$, $N(3,1)$, and $N(4,1)$ are independently distributed binomial random variables, it is possible to express the distribution of the data as follows:
\begin{align}
P(\mathbf{G}=\mathbf{g}\mid \mathbf{t},p) &= \sum_{\ell=0}^{t(1)} P\Big(N(1,2)=\ell \mid \mathbf{t},p\Big) \nonumber \\
&\quad\;\ \times P\Big(N(2,2)=g(2)-\ell \mid \mathbf{t},p\Big) \nonumber \\
&\quad\;\ \times P\Big(N(3,1)=t(1)+t(3)-g(4)-\ell \mid \mathbf{t},p\Big) \nonumber \\
&\quad\;\ \times P\Big(N(4,1)=g(1)+g(4)+\ell-t(1)-t(3) \mid \mathbf{t},p\Big) \nonumber \\
&\quad\;\ \times \mathbf{1}\left\{\sum_{j=1}^4 g(j) = \sum_{i=1}^{4}t(i)\right\} \nonumber \\
&= \sum_{\ell=0}^{t(1)} \mathrm{binom}\Big(\ell,t(1),p\Big) \nonumber \\
&\quad\;\ \times \mathrm{binom}\Big(g(2)-\ell,t(2),p\Big) \nonumber \\
&\quad\;\ \times \mathrm{binom}\Big(t(1)+t(3)-g(4)-\ell,t(3),p\Big) \nonumber \\
&\quad\;\ \times \mathrm{binom}\Big(g(1)+g(4)+\ell-t(1)-t(3),t(4),p\Big) \nonumber \\
&\quad\;\ \times \mathbf{1}\left\{\sum_{j=1}^{4}g(j)=\sum_{i=1}^{4}t(i)\right\} \nonumber
\end{align}
where $\mathbf{1}\{\cdot\}$ is the indicator function and $\mathrm{binom}(\cdot)$ is the binomial probability mass function,
\begin{align*}
\mathrm{binom}(k,r,p) = P(K=k\mid r,p) = \binom{r}{k} p^k (1-p)^{r-k}.
\end{align*}

\section{Estimation}\label{sec:estimation}

\subsection{Maximum Likelihood Estimator}

Using the probability of the realized data vector $\mathbf{g}$ in terms of the vector of potential outcome types $\mathbf{t}$ and the intended fraction of participants in the intervention group $p$, I can express the likelihood of the vector of potential outcome types $\mathbf{t}$ in terms of the realized data vector $\mathbf{g}$ and the intended fraction of participants in the intervention group $p$ as follows:
\begin{align*}
\mathcal{L}(\mathbf{t} \mid \mathbf{g},p) &= P(\mathbf{G}=\mathbf{g} \mid \mathbf{t},p) 
\end{align*}

\noindent I use the likelihood to specify the following maximum likelihood estimator \begin{align*}
\max_{\mathbf{t} \in \mathbb{N}_{0}^4} \; \mathcal{L}(\mathbf{t} \mid \mathbf{g},p) \nonumber 
\end{align*} 
where $\mathbf{t} \in \mathbb{N}_{0}^{4}$ indicates that the four elements of the vector $\mathbf{t}$ must belong to the set that includes the natural numbers and zero, which follows because they represent the counts of participants of each potential outcome type $i$.

As specified, the maximum likelihood estimator is a nonlinear integer programming problem.  It is possible that an analytical solution to this problem exists, but I have not be able to attain it using the software package Mathematica \citep{math2018}.\footnote{Mathematica's Maximize command is unable to maximize this expression and simply returns the input.  Using real interpolations of the factorial function, I can obtain partial derivatives of the likelihood function, but Mathematica's Solve command cannot find a critical point, returning the error ``Solve::nsmet: This system cannot be solved with the methods available to Solve."}  I have also not been able to attain a numerical solution using Mathematica.\footnote{Mathematica's NSolve function hangs indefinitely, while FindRoot results in errors due to limitations of machine-precision numbers.} It is possible that this problem could be solved numerically with another software package via a grid search, but I have not been able to find a solution through such an approach because the size of the grid is large for experiments of reasonable size, and evaluation of the objective function produces numbers that are the same within machine precision.  Software packages such as BARON \citep{baron}, can solve some types of nonlinear integer programming problems numerically when grid search algorithms are intractable.  However, BARON cannot solve this problem directly because it does not allow binomial coefficient terms to be specified within the objective function, nor does it allow a choice variable to appear as the upper limit of summation.\footnote{BARON limits the structure of the objective and constraint functions.  These functions must be expressed in terms of a constant number of algebraic operations.  While binomial coefficient terms can be specified as a string of products, the number of terms in the product depends on the input values.  Similarly, varying the upper limit of summation results in a varying number of summation terms.} I see the application of computational advances to this nonlinear programming problem as a promising avenue for future work. 

\subsection{Least Squares Estimator}

Since the maximum likelihood estimator is not yet computationally tractable via a variety of approaches, I propose an alternative least squares estimator that is. Intuitively, the maximum likelihood estimator maximizes a function that depends on the known functional form of the randomization error within the experiment.  The least squares estimator minimizes a function of the randomization error within the experiment using moments of its known functional form.

To be precise, define $\varepsilon(i)$, the randomization error within each potential outcome type $i$, as the difference between the actual and expected numbers of participants of type $i$ randomized into the intervention group, in terms of the intended fraction of participants in the intervention group $p$, the number of type $i$ participants randomized into the intervention group $N(i,1) + N(i,2)$, and the total number of type $i$ participants $t(i)$: 

\begin{equation*}
\varepsilon(i) = N(i,1) + N(i,2) - p*t(i).
\end{equation*}

\noindent From the model, we know that $N(i,1) + N(i,2)$ is distributed according to a binomial distribution with parameters $t(i)$ and $p$.  The mean of a binomial random variable is equal to the product of the parameters of the binomial distribution.  Therefore, it follows that the mean of the randomization error $\varepsilon(i)$ within each potential outcome type $i$ is zero: $E[\varepsilon(i)]=p*t(i)-p*t(i)=0$.

Furthermore, we can characterize the variance of the randomization error within each potential outcome type $i$. The variance of a binomial random variable is equal to the product of the probability of success, the probability of failure, and the number of trials.  Subtracting a constant from a binomial random variable does not change its variance.  Therefore, it follows that $Var[\varepsilon(i)]=p(1-p)*t(i)$.  Note that the variance of the randomization error increases as the number of participants increases.  A randomization error of 10 would be unlikely in a sample of 10, but it would be much more likely in a sample of 1000.  

Just as is it possible to characterize the mean and variance of the randomization error within each potential outcome type, it is also possible to characterize the mean and variance of the randomization error in any sample that combines all participants from two or more potential outcome types, including the experiment as a whole.  Let $I\subseteq\{1,2,3,4\}$ represent any subset of the potential outcome types.  Using this notation, we can express the the randomization error in any sample formed by combining all participants from every potential outcome type included in $I$ as follows:
\begin{align}
\varepsilon(I) = \sum_{i\in I}\left[N(i,1)+N(i,2)\right] - p\sum_{i\in I}t(i). \label{eq:err}
\end{align}
It is trivial show that $\varepsilon(I)$ has mean zero and variance equal to ${p(1-p)\sum_{i\in I}t(i)}$, which is proportional to the total number of participants $\sum_{i\in I}t(i)$.  

Using the mean and the variance of the randomization error in each subset of potential outcome types $I$, I construct an objective function $\mathcal{S}$.  This objective function represents the sum of the squared randomization error, weighted by the inverse of the variance of the randomization error, across all subsets of potential outcome types $I$ within the experiment: 
\begin{align*}
\mathcal{S}(\mathbf{N}\mid p) &= \sum_{I} \frac{1}{p(1-p)\sum_{i\in I} t(i)}\varepsilon(I)^2 \\
&=\sum_{I} \frac{1}{p(1-p)\sum_{i\in I} \{N(i,1)+N(i,2)+N(i,3)+N(i,4)\}} \\
&\quad\;\ \times  \left\{\sum_{i\in I}\left[N(i,1)+N(i,2)\right] - p\sum_{i\in I}t(i)\right\}^2,
\end{align*}
where I have shown that $\mathcal{S}$ is a function of the elements $N(i,j)$ of the matrix $\mathbf{N}$ by substituting for $\varepsilon(I)$ via (\ref{eq:err}) and substituting by for $t(i)=N(i,1)+N(i,2)+N(i,3)+N(i,4)$.  Intuitively, because the variance of the randomization error should be larger in larger subsamples, the objective function imposes a larger penalty for randomization error in smaller subsamples.

The least squares estimator minimizes the objective function $\mathcal{S}$ with respect to $\mathbf{N}$, subject to constraints that impose that $\mathbf{N}$ must give rise to the observed data vector $\mathbf{G}$ and that some elements of $\mathbf{N}$ must be equal to zero by deductive reasoning based on the shaded cells of the matrix in Figure~\ref{fig:expMatrix}: 
\begin{align*}
\min_{\mathbf{N} \in \mathbb{N}_{0}^{4\times 4}} &\; S(\mathbf{N}\mid p) \label{pr:ss}\\
\textrm{s.t.}
&\; \sum_{i=1}^{4} N(i,j) = g(j) \; \mathrm{for}\  j=1,2,3,4 \nonumber \\
&\; N(1,1)=N(1,3)=N(2,1)=N(2,4)=N(3,2)=N(3,3)=N(4,2)=N(4,4)=0, \nonumber
\end{align*}
where $\mathbf{N} \in \mathbb{N}_{0}^{4\times 4}$ indicates that each element $N(i,j)$ of the 4 by 4 matrix $\mathbf{N}$ must belong to the set that includes the natural numbers and zero. I denote the estimate of each $N(i,j)$ with $\hat{N(i,j)}$, and I construct the least squares estimate of each potential outcome type $\hat{t}(i)$ as follows:   
\begin{equation*}
\hat{t}(i) = \sum_{j=1}^4 \hat{N}(i,j) \qquad \forall i=1,2,3,4.
\end{equation*}
Unlike the maximum likelihood estimator, the least squares estimator involves a nonlinear programming problem of a form that can be estimated numerically in BARON.

\section{Empirical Application} \label{sec:empirical}
\subsection{The PROWESS Clinical Trial}	

To apply my model to the PROWESS clinical trial, I use aggregated data reported in \citet{bernard2001}, which allow me to replicate the main reduced form result from \citet{bernard2001} exactly.  The main reduced form result is based on the comparison of 28-day mortality ($Y=1$ if dead, $Y=0$ if alive) for 850 participants assigned to the intervention group and 840 participants assigned to the control group ($Z=1$ if intervention, $Z=0$ if control). \citet{bernard2001} report that participants were randomized into the intervention and control groups in a 1:1 manner, so I set the intended fraction of participants in the intervention group to $p=1/2$.  

By day 28, 210 participants in the intervention group had died and 259 participants in the control group had died.   Therefore, as I report across the columns of the matrix in Figure~\ref{fig:pwsRf}, $g(1) = 210$ participants were dead in the intervention group, $g(2) = 640$ participants were alive in the intervention group, $g(3) = 259$ participants were dead in the control group, and $g(4) = 581$ participants were alive in the control group.  The reduced form based on these statistics shows that assignment to the intervention group decreased mortality by 6 percentage points (0.06 is approximately equal to 210/850 - 259/840), which is statistically different from 0 at the 0.5\% level.  An alternative reduced form reported in \citet{bernard2001} shows that assignment to the intervention group increased the incidence of serious bleeding by 1.5 percentage points, which is statistically different from 0 at the 6\% level, suggesting a potential mechanism through which participants could be killed within the trial.    

\subsection{Results} \label{subsec:empEst}

\begin{figure}[!hbt]
	\caption{Matrix that Relates Potential Outcome Types and Observed Outcome Groups  \\
	Estimates from the PROWESS Trial}
	\centering
	\includegraphics[width=1\linewidth]{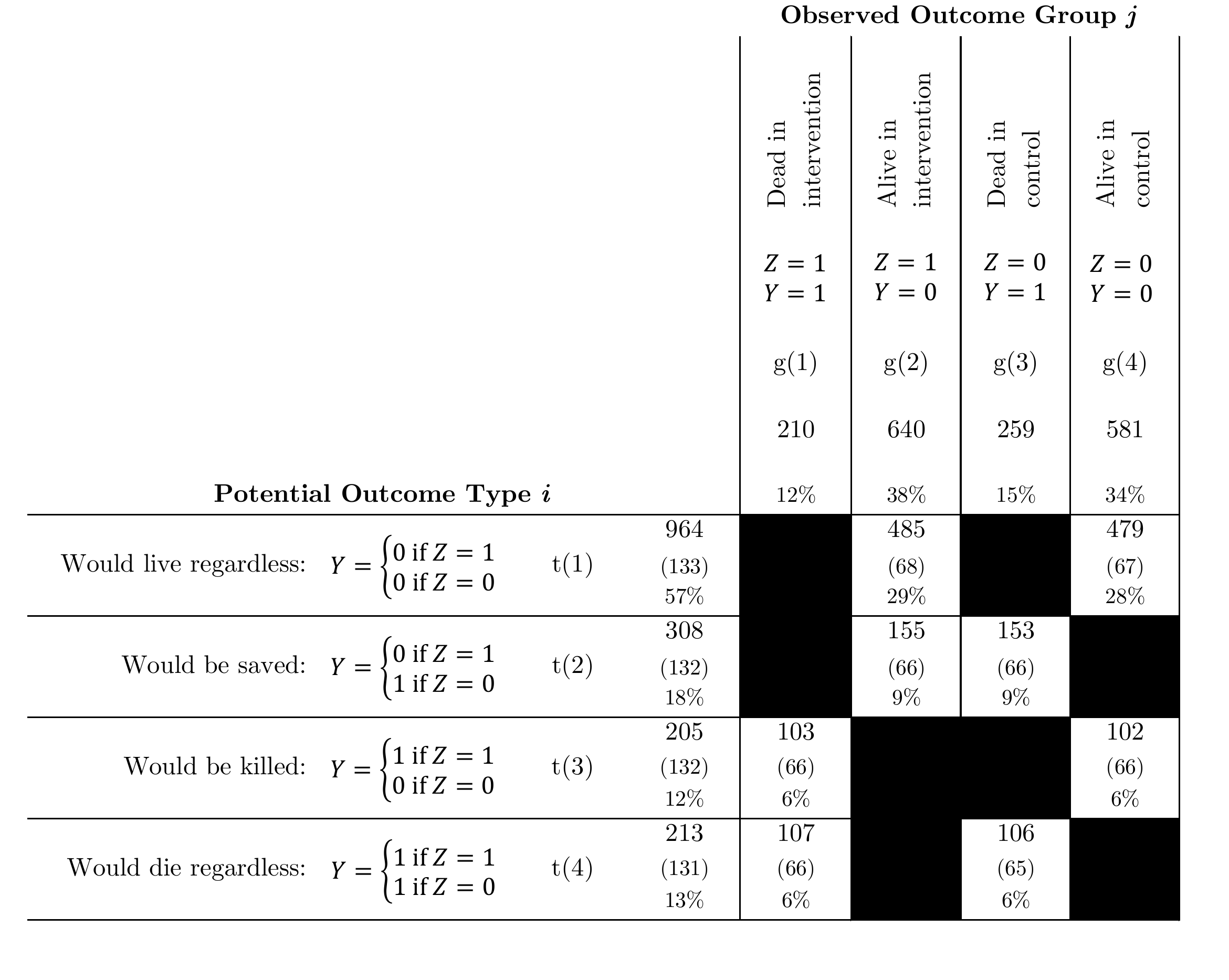}
	\label{fig:pwsRf}
	\vspace{-30pt}
	\begin{minipage}{1\linewidth}
	\scriptsize
	\emph{Note}.  Bootstrapped standard errors are reported in parentheses. Percentages represent percentages of the full sample.  They do not necessarily sum to 100\% due to rounding. $Y$ represents 28-day mortality, and $Z$ represents assignment to the intervention group in the PROWESS trial. $N(i,j)$, the number of participants of potential outcome type $i$ in observed outcome group $j$, must be equal to zero in all shaded cells. The observed outcome group vector $\mathbf{g}$ and the intended fraction of participants in the intervention group $p=1/2$  are taken from the PROWESS trial. The values of the matrix $\mathbf{N}$ and the potential outcome type vector $\mathbf{t}$ are estimates obtained via the least squares estimator, estimated using the BARON optimization package within MATLAB.  
	\end{minipage}

\end{figure}

I report the results that I obtained by applying the least squares estimator to the PROWESS trial in the rows of the matrix in Figure~\ref{fig:pwsRf}.  I obtained the estimates by using the BARON optimization package \citep{baron} within MATLAB \citep{matlab2016}.  The estimates show that $t(1)=964$ participants would live regardless, $t(2)=308$ participants would be saved, $t(3)=205$ participants would be killed, and $t(4)=213$ participants would die regardless.  

As a fraction of the full sample of 1690 participants, 57\% would live regardless, 18\% would be saved, 12\% would be killed, and 13\% would die regardless. Note that the absolute difference between the number of participants saved and killed is 6\%, which is mechanically equal to the reduced form estimate.  However, the least squares estimates provide context for the reduced form estimate by showing that 70\% of participants would not be affected by assignment to the intervention and that 2 participants would be killed for every 3 participants saved (2/3=12\%/18\%).    Because of the random assignment, only about half of the patients who would be killed were assigned to the intervention group. One element of the intermediate output of my least squares estimator, $n(3,1)$, reported in the first column of the third row of the matrix in Figure~\ref{fig:pwsRf}, shows that 103 trial participants were killed.  

To assess the statistical significance of these estimates, I report standard errors that I obtained as the standard deviation of the estimates that I obtained via 1000 bootstrap iterations. The standard errors are very similar across the potential outcome types, so statistical significance tends to increase with the number of participants of each potential outcome type. The number of participants who would live regardless is the largest ($t(1)=964$), and it is statistically different from zero at conventional levels.  The number of participants who would be saved ($t(2)=308$) is also statistically different from zero at conventional levels.  

\section{Monte Carlo}\label{sec:montecarlo}

\subsection{Design}
To examine the performance of the least squares estimator, I perform a Monte Carlo simulation.  In each simulated experiment $m$, I set the true vector of the number of participants of each potential outcome type $\mathbf{t}$ to the estimated vector from the PROWESS trial.  Using the intended fraction of participants in the intervention group $p=1/2$,  I randomly assign each participant to the intervention or control group.  I use each participant's potential outcome type $i$ to determine each participant's observed outcome group $j$.  For example, if I assign a participant who would live regardless ($i=1$) to the intervention group, then I recognize that the participant will live and thus be observed in outcome group $j=2$.  Aggregating across all participants within a simulated experiment, I determine the observed outcome group vector $\mathbf{g}$.  Using the observed outcome group vector $\mathbf{g}$ and the intended fraction of participants in the intervention group $p=1/2$, I apply the least squares estimator and store my estimates of the potential outcome type vector $\mathbf{t}$.  I run a total of $M=1000$ simulated experiments.  Across all experiments, I construct the mean bias and root mean square error (RMSE) for each element of the potential outcome type vector $\mathbf{t}$.

\subsection{Results}
 
\noindent I report the results of the Monte Carlo simulation in Table~\ref{tab:expMc}.  The results show that the mean bias and RMSE do not vary much across potential outcome types.  For each potential outcome type $i$, the mean bias in the total number of participants $t(i)$ is approximately 42 participants, which represents  2.5\% of the full sample of 1690 participants.  The RMSE is approximately 124 participants for each type $i$, which represents 7.3\% of the full sample of 1690 participants.

\begin{table}[!hbt]
	\centering	
	\caption{Monte Carlo Simulation Results}
	\includegraphics{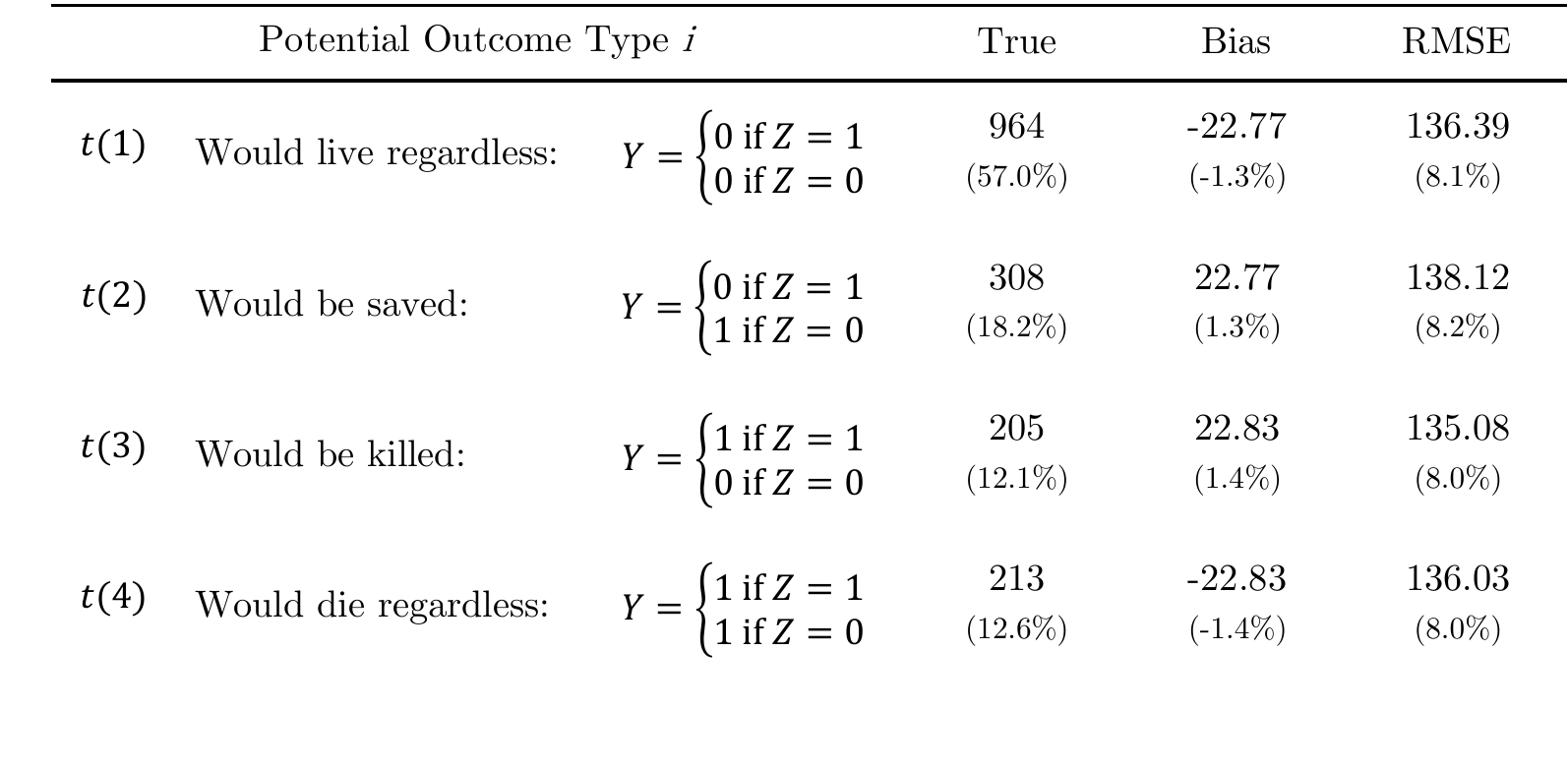} \\
	\label{tab:expMc}
	\vspace{-20pt}
	\begin{minipage}{1\linewidth}
		\scriptsize
		\begin{onehalfspace}
		\emph{Note}. The Monte Carlo simulation generates $M=1000$ simulated experiments in which the true potential outcome type vector $\mathbf{t}$ is equal to the estimated potential outcome type vector from the PROWESS trial. In each simulation, the observed outcome group vector $\mathbf{g}$ is calculated via the model, using the intended fraction of participants $p=1/2$ from the PROWESS trial.  Within each simulated experiment, $\mathbf{g}$ and $p=1/2$ are the inputs used to estimate the potential outcome type vector $\mathbf{t}$ with the least squares estimator, via the BARON optimization package within MATLAB.  Letting $t(i)$ be the number of participants of true potential outcome group $i$ and $\hat{t}_m(i)$ be the estimate of number of participants of true potential outcome group $i$ in simulation $m$, mean bias is $\frac{1}{M}\sum_m (\hat{t}_m(i)-t(i))$, and RMSE is $\left(\frac{1}{M}\sum_m (\hat{t}_m(i)-t(i))^2\right)^{1/2}$. Percentages represent percentages of the full sample. 
	\end{onehalfspace} 	  		
	\end{minipage}

\end{table}

\section{Conclusion}\label{sec:conclusion}

In this paper, I develop a model of a randomized experiment. Although I have focused on a randomized experiment, the model could also be adapted for application to any natural experiment or instrumental variable setting. When applied to the PROWESS clinical trial, the model and the associated least squares estimator that I develop allow me to estimate the number of participants who would live regardless, the number of participants who would be saved, the number of participants who would be killed, and the number of participants who would die regardless.  My estimates show that the intervention within the PROWESS trial killed two participants for every three it saved. These estimates have ethical implications.

\bibliographystyle{chicago}
\bibliography{experiment}

\begin{thebibliography}{}

\bibitem[\protect\citeauthoryear{Bernard, Vincent, Laterre, LaRosa, Dhainaut,
  Lopez-Rodriguez, Steingrub, Garber, Helterbrand, Ely, and Fisher}{Bernard
  et~al.}{2001}]{bernard2001}
Bernard, G.~R., J.-L. Vincent, P.-F. Laterre, S.~P. LaRosa, J.-F. Dhainaut,
  A.~Lopez-Rodriguez, J.~S. Steingrub, G.~E. Garber, J.~D. Helterbrand, E.~W.
  Ely, and C.~J. Fisher (2001).
\newblock Efficacy and safety of recombinant human activated protein c for
  severe sepsis.
\newblock {\em New England Journal of Medicine\/}~{\em 344\/}(10), 699--709.
\newblock PMID: 11236773.

\bibitem[\protect\citeauthoryear{Foot}{Foot}{1967}]{foot1967}
Foot, P. (1967).
\newblock The problem of abortion and the doctrine of double effect.
\newblock {\em Oxford Review\/}~{\em 5}, 5--15.

\bibitem[\protect\citeauthoryear{Holland}{Holland}{1986}]{holland1986}
Holland, P.~W. (1986).
\newblock Statistics and causal inference.
\newblock {\em Journal of the American statistical Association\/}~{\em
  81\/}(396), 945--960.

\bibitem[\protect\citeauthoryear{Rubin}{Rubin}{1974}]{rubin1974}
Rubin, D.~B. (1974).
\newblock Estimating causal effects of treatments in randomized and
  nonrandomized studies.
\newblock {\em Journal of educational Psychology\/}~{\em 66\/}(5), 688.

\bibitem[\protect\citeauthoryear{Rubin}{Rubin}{1977}]{rubin1977}
Rubin, D.~B. (1977).
\newblock Assignment to treatment group on the basis of a covariate.
\newblock {\em Journal of Educational and Behavioral statistics\/}~{\em
  2\/}(1), 1--26.

\bibitem[\protect\citeauthoryear{Sahinidis}{Sahinidis}{2018}]{baron}
Sahinidis, N.~V. (2018).
\newblock {\em {BARON 18.8.23: Global Optimization of Mixed-Integer Nonlinear
  Programs, {\em User's Manual}}}.

\bibitem[\protect\citeauthoryear{Siegel}{Siegel}{2002}]{siegel2002}
Siegel, J.~P. (2002).
\newblock Assessing the use of activated protein c in the treatment of severe
  sepsis.
\newblock {\em The New England journal of medicine\/}~{\em 347\/}(13),
  1030--1034.

\bibitem[\protect\citeauthoryear{{The Mathworks{,} Inc.}}{{The Mathworks{,}
  Inc.}}{2016}]{matlab2016}
{The Mathworks{,} Inc.} (2016).
\newblock Matlab, {V}ersion r2016a.
\newblock Natick, Massachusetts, United States.

\bibitem[\protect\citeauthoryear{Thomson}{Thomson}{1985}]{thomson1985}
Thomson, J.~J. (1985).
\newblock The trolley problem.
\newblock {\em Yale Law Journal\/}~{\em 94\/}(1395), 1395--1415.

\bibitem[\protect\citeauthoryear{{US Food and Drug Administration and
  others}}{{US Food and Drug Administration and others}}{2011}]{fda2011}
{US Food and Drug Administration and others} (2011).
\newblock {FDA} drug safety communication: voluntary market withdrawal of
  xigris due to failure to show a survival benefit.
\newblock {\em US Food and Drug Administration, Washington, DC\/}.

\bibitem[\protect\citeauthoryear{Warren, Suffredini, Eichacker, and
  Munford}{Warren et~al.}{2002}]{warren2002}
Warren, H.~S., A.~F. Suffredini, P.~Q. Eichacker, and R.~S. Munford (2002).
\newblock Risks and benefits of activated protein c treatment for severe
  sepsis.
\newblock {\em The New England journal of medicine\/}~{\em 347\/}(13),
  1027--1030.

\bibitem[\protect\citeauthoryear{{Wolfram Research{,} Inc.}}{{Wolfram
  Research{,} Inc.}}{2018}]{math2018}
{Wolfram Research{,} Inc.} (2018).
\newblock Mathematica, {V}ersion 11.3.
\newblock Champaign, IL.

\end{thebibliography}

\end{document}